44

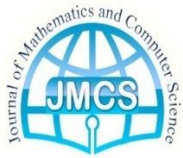
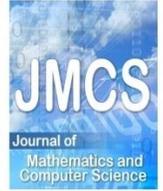

Contents list available at JMCS

**Journal of Mathematics and Computer Science**

Journal Homepage: www.tjmcs.com

# Experimental Estimation of Number of Clusters Based on Cluster Quality

G. Hannah Grace, Kalyani Desikan

*Department of Mathematics, School of Advanced Sciences, VIT University, Chennai 600127, India*

hannahgrace.g@vit.ac.in, kalyanidesikan@vit.ac.in

*Abstract*

Text Clustering is a text mining technique which divides the given set of text documents into significant clusters.  It is used for organizing a huge number of text documents into a well-organized form.  In the majority of the clustering algorithms, the number of clusters must be specified apriori, which is a drawback of these algorithms.  The aim of this paper is to show experimentally how to determine the number of clusters based on cluster quality. Since partitional clustering algorithms are well-suited for clustering large document datasets, we have confined our analysis to a partitional clustering algorithm.

## 1. Introduction

Text clustering is used to understand the structure and content of unknown text sets as well as to give new perspectives on familiar ones. The main aim of text clustering is to minimize intra-cluster distance between documents, and maximize the inter-cluster distance using an appropriate distance or similarity measure.  In this paper, we have experimentally found the maximum number of clusters into which the document set is to be partitioned based on cluster quality. We have used the CLUTO clustering tool to evaluate the entropy and purity of clusters, by varying the number of clusters.

Our paper is organized as follows.  Section 2 introduces the major types of clustering i.e., partitional and hierarchical clustering. Section 3 to 5 explains the preliminaries like similarity measures, criterion functions and cluster qualities like entropy and purity. Section 6 describes the data sets used in our experiment and description about the clustering tool CLUTO.  Section 7 explains the experiments performed , their results and their interpretation. Section 8 concludes the results to show that the Repeated bisection method performs the best with respect to clustering time. Also the quality of clusters is analyzed to find the optimal number of clusters. Also, we have shown that among all the criterion functions, the I2 criterion performs the best with respect to clustering time.

## 2. Types of clustering

Clustering methods are classified based on the cluster structure they produce. Most clustering methods require an object representation and also a distance or similarity measure between objects. The major clustering methods can be classified as Partitioning methods and Hierarchical methods.

### 2.1. Partitioning clustering

Given $D$, a data set of $n$ objects, and $k$, the number of clusters to form, a partitioning algorithm organizes the objects into $k$ partitions ($k \leq n$), where each partition represents a cluster. The clusters are formed to optimize an objective partitioning criterion, such as a dissimilarity function based on distance, so that the objects within a cluster are 'similar' whereas the objects belonging to different clusters are 'dissimilar'. The commonly used partitioning methods are k-means, k-medoids, and their variations [1].

### 2.2. Hierarchical clustering

A hierarchical clustering method groups data objects into a tree of clusters. The Hierarchical clustering algorithms, are classified as agglomerative or divisive, depending on whether the hierarchical decomposition is formed in a bottom-up (merging) or top-down (splitting) manner. The main drawback of a pure hierarchical clustering method is that once a merge or split decision is taken and if it turns out to be a poor choice, this method cannot backtrack and correct it [1].

Agglomerative methods [2], when used for document clustering, starts with an initial clustering where each document is considered to be a separate cluster. The closest clusters, using a given inter cluster similarity measure, are then merged repeatedly until only one cluster or a predefined number of clusters remain.

Divisive clustering algorithms start with a single cluster containing all the documents. It is then repeatedly divided into clusters until all documents are contained in their own cluster or a predefined number of clusters [3].

In recent years, various researchers have recognized that partitional clustering algorithms are well-suited for clustering large document datasets because of their relatively low computational requirements.

## 3. Similarity measure

A similarity measure is a function which computes the degree of similarity between pairs of texts. A similarity measure can represent the similarity between two documents, two queries, or one document and one query.

Similarity measure assigns a real number between 0 and 1 to a pair of documents, depending upon the degree of similarity between them. A value of zero signifies that the documents are completely dissimilar while a value of one indicates that the documents are identical.

Traditionally, vector-based models have been used for computing the document similarity. The most commonly used proximity measure in vector-based models is the Euclidean distance given by the general formula $D_{ij} = \left( \sum_{k=1}^{d} |x_{ik} - x_{jk}|^2 \right)^{1/2}$ where d is the dimensionality of the data object and $x_{ik}$ and $x_{jk}$ are the $k^{th}$ components of the $i^{th}$ and $j^{th}$ objects $x_i$ and $x_j$. $D_{ij}$ is the proximity between points $x_i$ and $x_j$.

Some of the distance measures used in text clustering are Euclidean Distance, squared Euclidean Distance, Normalized squared Euclidean Distance, Manhattan Distance, Edit Distance, Hamming Distance; while the similarity measures include cosine similarity, Pearson correlation coefficient, Jaccard similarity, Dice coefficient, overlap coefficient, asymmetric similarity and Averaged Kullback-Leibler Divergence.

## 4. Criterion function

A key characteristic of many partitional clustering algorithms is that they use a global criterion function whose optimization drives the entire clustering process [5]. Some of the criterion functions that are used for finding the clusters are I1, I2, E1, G1, G1p, H1 and H2.
The descriptions for the criterion functions are as follows:
1. *I1* criterion function – This function tries to maximize the intra cluster similarity between the elements of a cluster.
2. *I2* criterion function – This function also tries to maximize the intra cluster similarity between the elements of a cluster. The only difference between I1 and I2 is that while calculating I2 we must take the square root of the function.
3. *E1* criterion function – This function divides the intra-cluster similarity with inter cluster similarity
4. *G1* criterion function – This function similar to E1 except that there is no square root in the denominator.
5. *G1p* criterion function – This function is also similar to *E1* except that we have $n_1^2$ and that there is no root in the denominator.
6. *H1* criterion function - This is a hybrid function to maximize *I1/E1*.
7. *H2* criterion function – This is a hybrid function trying to maximize *I2/E1*.

The mathematical formulae for these functions are given in the following table.

**Table 1**

| Criterion function | Formula |
|---|---|
| *I1* | maximize $\sum_{i=1}^{k} \frac{1}{n_i} \left( \sum_{u,v \in S_i} sim(u,v) \right)$     (1) |
| *I2* | maximize $\sum_{i=1}^{k} \sqrt{\sum_{u,v \in S_i} sim(u,v)}$     (2) |

| | | |
|---|---|---|
| E1 | minimize $\sum_{i=1}^{k} n_i \dfrac{\sum_{u \in S_i, v \in S} sim(u,v)}{\sqrt{\sum_{u,v \in S_i} sim(u,v)}}$ | (3) |
| G1 | minimize $\sum_{i=1}^{k} \dfrac{\sum_{u \in S_i, v \in S} sim(u,v)}{\sum_{u,v \in S_i} sim(u,v)}$ | (4) |
| G1p | minimize $\sum_{i=1}^{k} n_i^2 \dfrac{\sum_{u \in S_i, v \in S} sim(u,v)}{\sum_{u,v \in S_i} sim(u,v)}$ | (5) |
| H1 | maximize $\dfrac{I1}{E1}$ | (6) |
| H2 | maximize $\dfrac{I2}{E1}$ | (7) |

**Table 1** gives the formulae for calculating the clustering criterion functions. The notation in these formulae are as follows: k is the total number of clusters, S is the total number of documents to be clustered, $S_i$ is the set of documents assigned to the $i^{th}$ cluster, $n_i$ is the number of documents in the $i^{th}$ cluster, u and v represent two documents and sim(u,v) is the similarity measure between these two documents.

These functions optimize the intra-cluster similarity, inter-cluster dissimilarity, and their combinations and represent some of the most widely-used criterion functions for document clustering.

Theoretical analysis of the criterion functions shows that their relative performance depends on the (i) degree to which they can correctly operate when the clusters are of different tightness, and (ii) degree to which they can lead to reasonably balanced clusters.

## 5. Cluster Quality

A cluster quality measure is a function that maps pairs of the form (dataset, clustering) to some ordered set (say, the set of non-negative real numbers), so that these values reflect how 'good' or 'cogent' that clustering is. Cluster quality measures may also be used to identify an ideal clustering method by comparing the different clustering solutions obtained when different clustering methods/parameters are employed over the same data set (e.g., comparing the results of a given clustering paradigm over different choices of clustering parameters, such as the number of clusters).

For the evaluation of cluster quality, we use two different measures: Entropy and Purity [6] [7]. These are standard measures that help us to ascertain the cluster quality. Entropy measures how the various semantic classes are distributed within each cluster. Given a particular cluster $S_r$ of size $n_r$, the entropy of this cluster is defined as

$$E(S_r) = -\frac{1}{\log q} \sum_{i=1}^{q} \frac{n_r^i}{n_r} \log \frac{n_r^i}{n_r}$$

where q is the number of classes in the dataset and $n_r^i$ is the number of documents of the i[th] class that are assigned to the r[th] cluster.  The entropy of the entire clustering solution is then the sum of the individual cluster entropies weighted according to the cluster size:

$$Entropy = \sum_{r=1}^{k} \frac{n_r}{n} E(S_r)$$

where n is the total number of documents.
Smaller entropy values indicate better clustering solutions.
Using the same Mathematical notation, the purity of a cluster is defined as [7]

$$Pu(S_r) = \frac{1}{n_r} \max n_r^i$$

The purity of the clustering solution is the weighted sum of the individual cluster purities

$$purity = \sum_{r=1}^{k} \frac{n_r}{n} Pu(S_r)$$

Larger purity values indicate better clustering solutions.
Entropy is a more comprehensive measure than purity because rather than just considering the number of documents, it considers the overall distribution of all the classes in a given cluster [8]. For an ideal cluster with documents from only a single class, the entropy of the cluster will be 0. In general, the smaller the entropy value, the better the quality of the cluster.

## 6. Experimental setup

### 6.1 Data sets used

For our experimental study, we have used a well known benchmark dataset used in text mining, the Classic collection [9].  This dataset consists of 4 different document collections: CACM, CISI, CRAN, and MED. The composition of the collection of classic data set is as follows:
- CACM: 3204 documents
- CISI: 1460 documents
- CRAN: 1398 documents
- MED: 1033 documents

This is also referred to as classic4 dataset.

### 6.2. CLUTO

The clustering tool we have used for our experimental study is CLUTO [9]. CLUTO can be used for clustering low and high dimensional datasets and for analyzing the characteristics of the various clusters. CLUTO can operate on very large set of documents as well as number of dimensions.
CLUTO performs Repeated Bisections, Repeated Bisections by k-way refinement, Direct k-way clustering, Agglomerative clustering, Graph partitioning based clustering and partitional based Agglomerative clustering.

CLUTO can be used to cluster documents based on the following similarity measures: cosine similarity, correlation coefficient, Euclidean distance and extended Jaccard coefficient. CLUTO provides cluster quality details such as Entropy and Purity.

## 7. Results and Interpretation

In order to identify the most appropriate Partitional method for our experimental study, we first observed the clustering time for the Repeated bisection, Direct and Agglomerative methods by varying the number of clusters for the classic dataset.

Figure 1 shows the clustering time for Repeated bisection, Direct and Agglomerative methods. We have used the I2 Criterion function for all the three methods.

From Figure 1, it is clear that the performance of the Repeated Bisection method is the best with respect to clustering time. As we increase the number of clusters, the time taken to form the clusters in the case of the Direct method increases rapidly. Though the time for clustering in the case of the Agglomerative method does not change as we vary the number of clusters, when compared to the Repeated bisection method the time taken is more. Hence, for our subsequent analysis, we have adopted the Repeated bisection method.

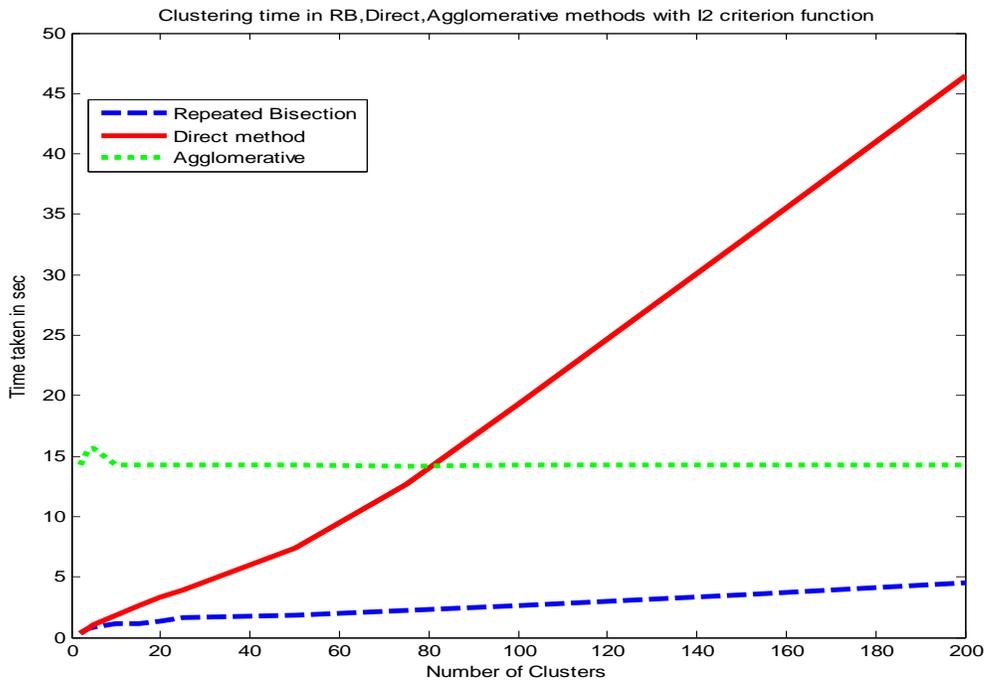

**Figure 1**

For identifying the maximum number of clusters, we have analysed and plotted the entropy and purity by applying the criterion functions I1, I2, E1 G1, G1p, H1, H2 to the Repeated bisection method as we increased the number of clusters.

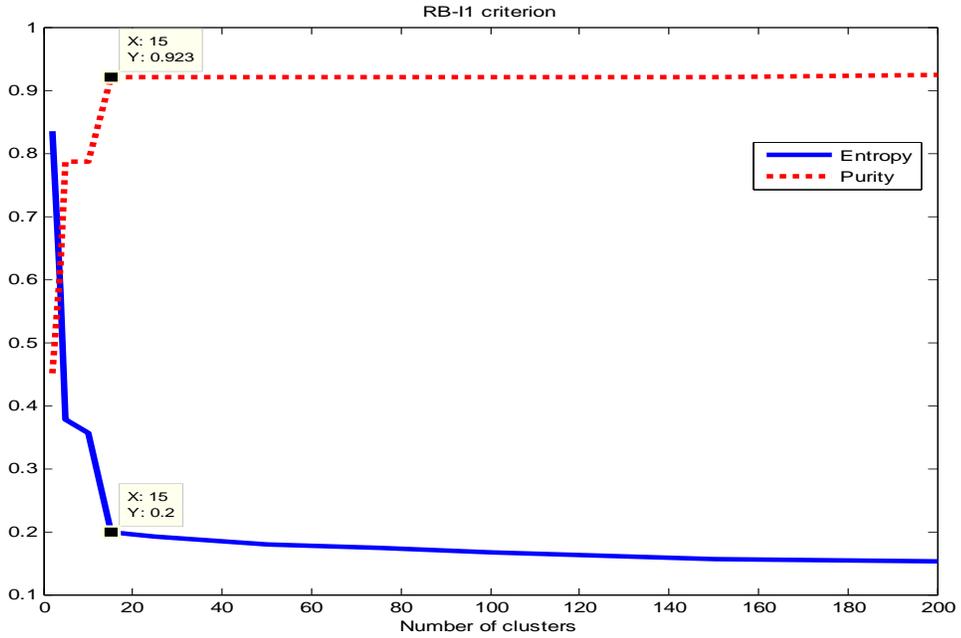

**Figure 2**

**Figure 2** gives the entropy and purity for *I1* criterion function. We notice that the entropy does not change much as we increase the number of clusters beyond 15. We can also see that entropy hovers around 0.195 beyond 15 clusters.

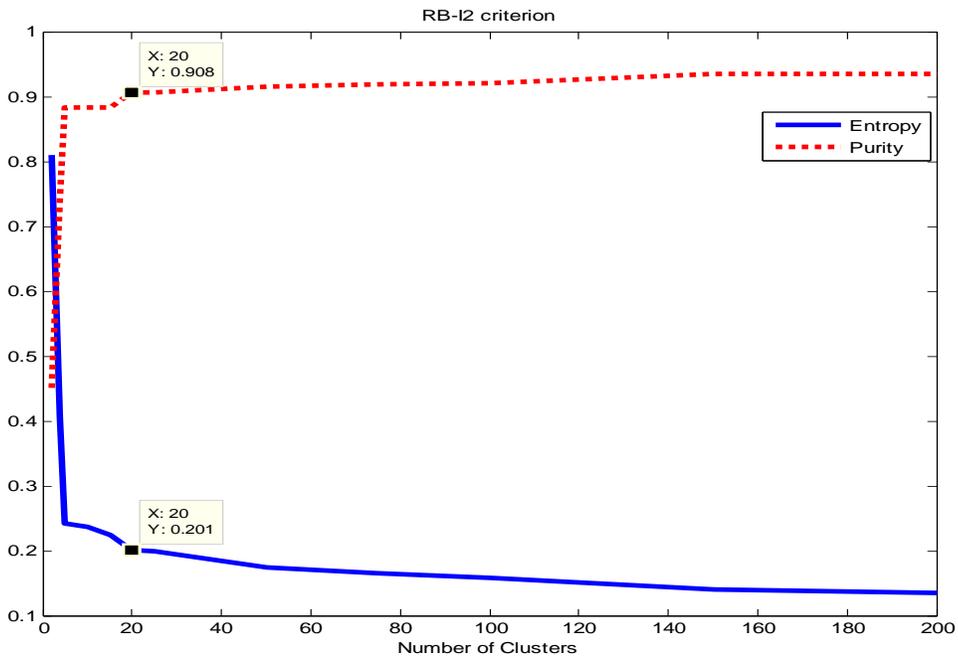

**Figure 3**

**Figure 3** shows the variation in entropy and purity for I2 criterion function. We notice that the entropy does not vary much even if we increase the number of clusters beyond 40. We can also see that purity hovers around 0.904 beyond 25 clusters.

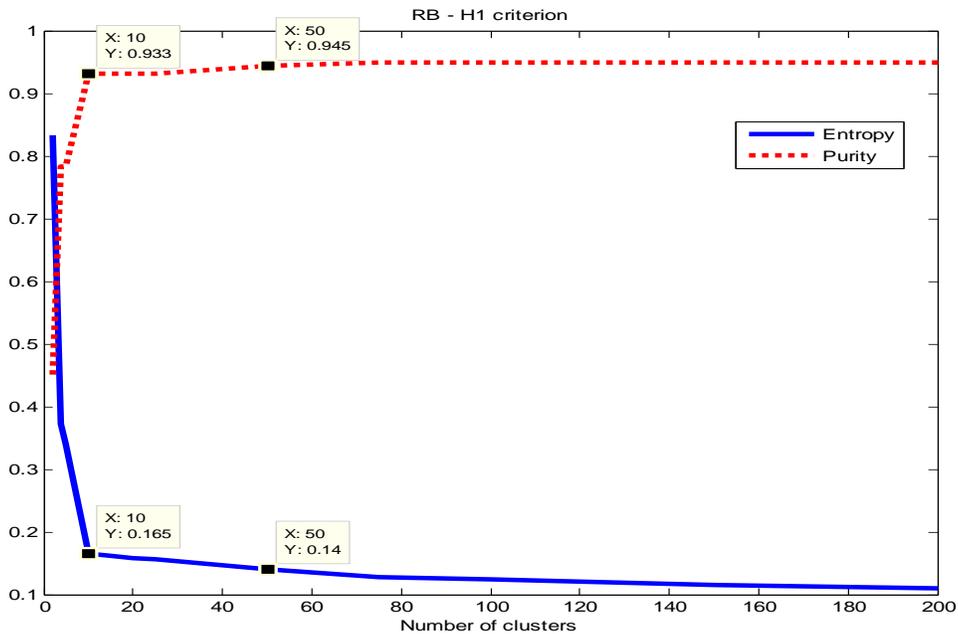

**Figure 4**

From **Figure 4**, we observe that the entropy varies very gradually as we increase the number of clusters beyond 50 . Also purity stabilises around 0.945 when the number of clusters reaches 50. Thereafter, the change in purity is not significant.

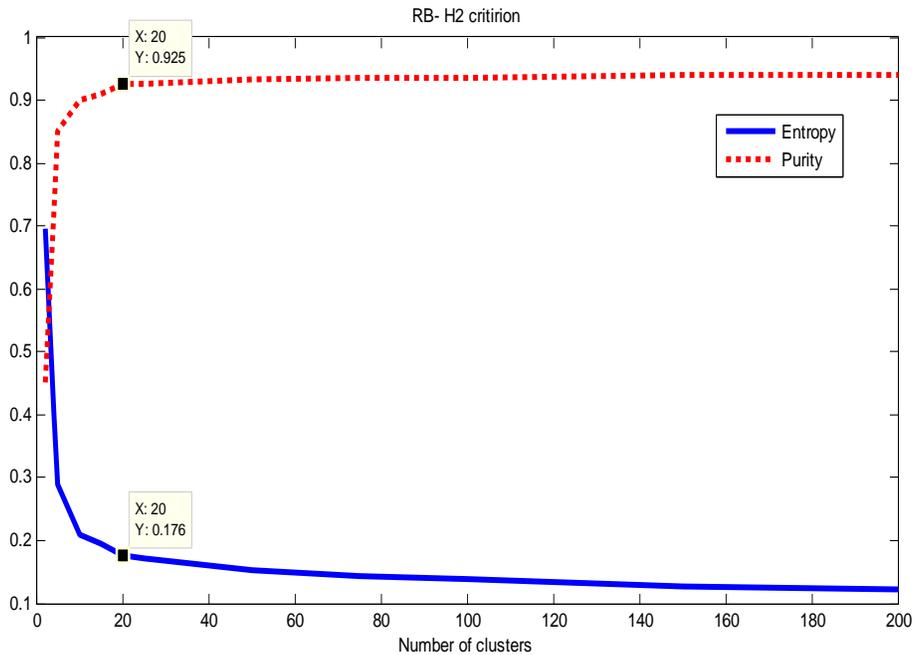

**Figure 5**

311

**Figure 5** depicts the entropy and purity as we vary the number of clusters for H2 criterion function. We notice that the entropy and purity do not change much beyond 40 clusters.

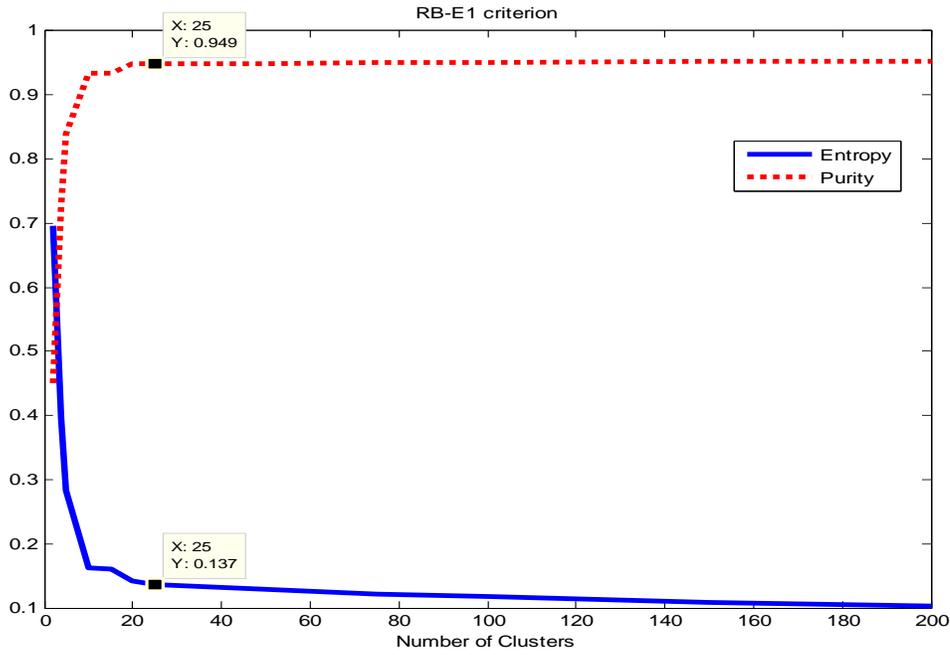

**Figure 6**

From **Figure 6**, it is clear that the entropy and purity values, when we apply the E1 criterion condition, stabilise when we reach 25 clusters.

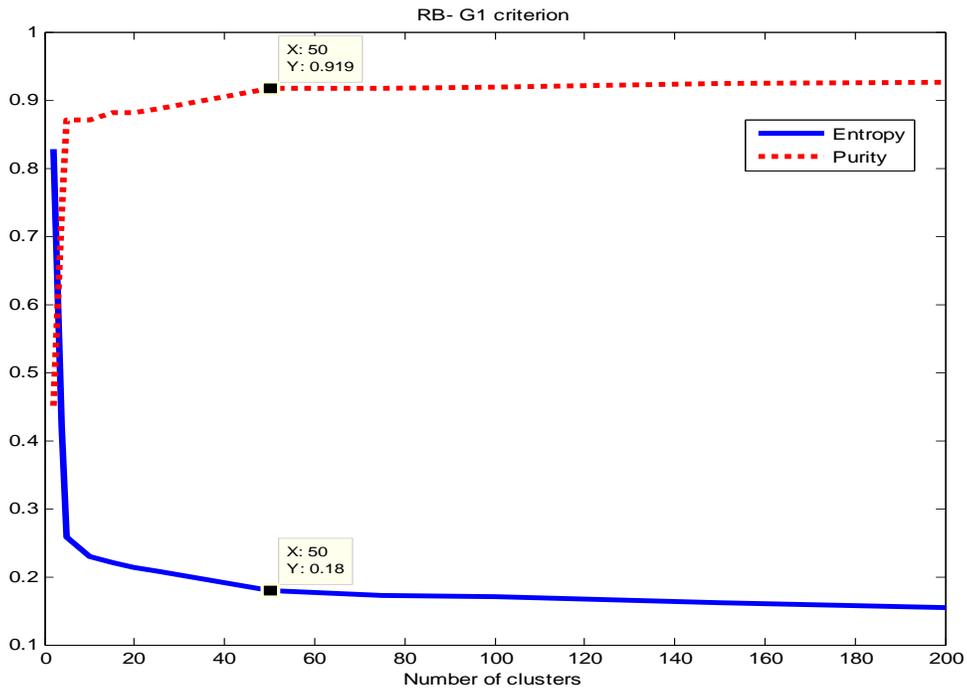

**Figure 7**

**Figure 7** gives the entropy and purity for G1 criterion condition. We notise that the entropy and purity do not change much beyond 50 clusters.

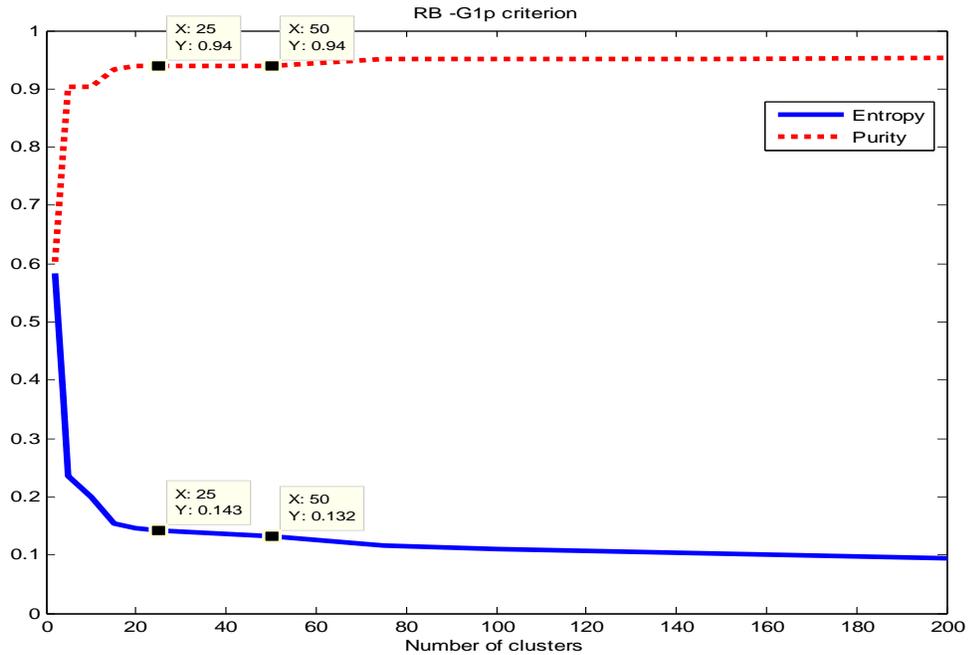

**Figure 8**

**Figure 8** shows the entropy and purity for the G1p criterion. We can see that the entropy becomes stable when the number of clusters is 50. Purity stabilizes at around 30 clusters. The following table gives the entropy and purity values marked in the **Figure (2-8)**

**Table 2**

| Figure | Number of clusters | Entropy | Purity |
|---|---|---|---|
| b | 15 | 0.2 | 0.923 |
| c | 20 | 0.201 | 0.908 |
| d | 10 | 0.155 | 0.933 |
|   | 50 | 0.14 | 0.945 |
| e | 20 | 0.176 | 0.925 |
| f | 25 | 0.137 | 0.949 |
| g | 50 | 0.180 | 0.919 |
| h | 25 | 0.143 | 0.940 |
|   | 50 | 0.132 | 0.940 |

For identifying the most appropriate criterion function, we found the time to cluster when different criterion functions were used in the Repeated bisection method. **Figure 9** gives the clustering time when we used the different criterion functions in the Repeated bisection method.

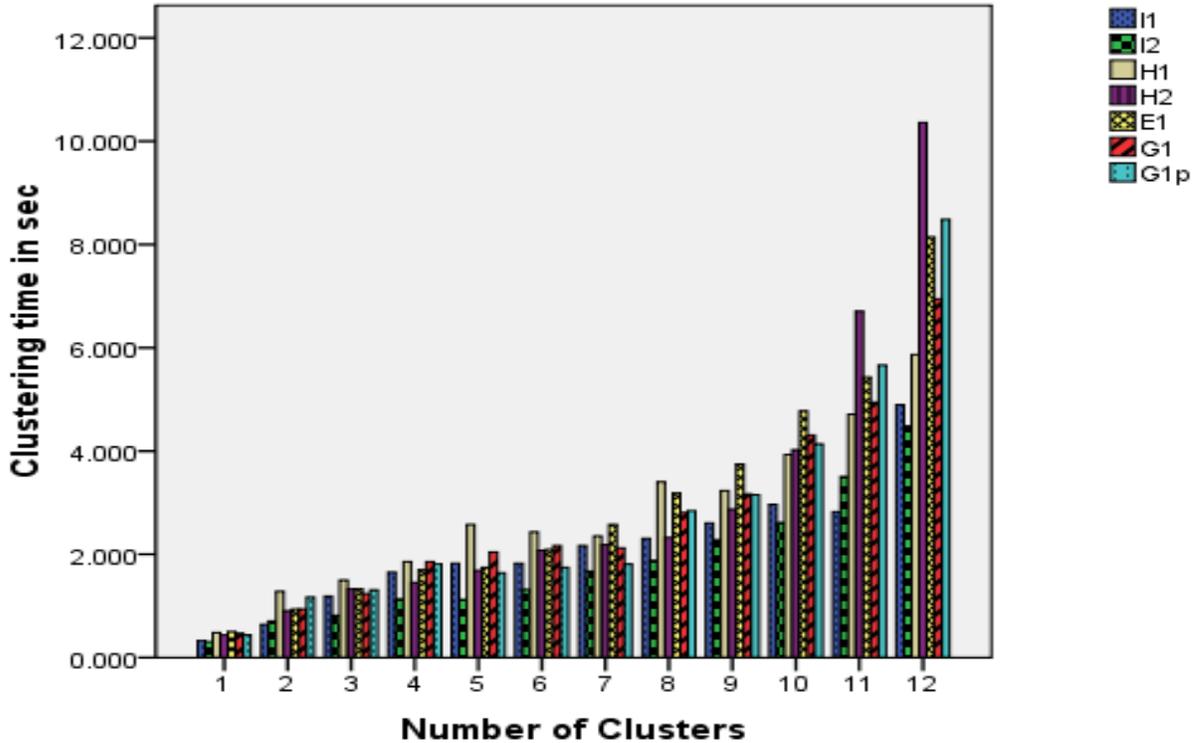

**Figure 9**

The following table gives the clustering time in seconds for the criterion functions depicted in **Figure 9**

**Table 3**

| Number of Clusters | Clustering time in sec for criterion functions | | | | | | |
|---|---|---|---|---|---|---|---|
| | I1 | I2 | H1 | H2 | E | G1 | G1p |
| 2 | 0.327 | 0.306 | 0.484 | 0.437 | 0.5 | 0.468 | 0.437 |
| 4 | 0.64 | 0.707 | 1.279 | 0.904 | 0.936 | 0.936 | 1.17 |
| 5 | 1.186 | 0.815 | 1.497 | 1.326 | 1.326 | 1.233 | 1.31 |
| 10 | 1.653 | 1.134 | 1.856 | 1.451 | 1.7 | 1.856 | 1.81 |
| 15 | 1.826 | 1.122 | 2.574 | 1.685 | 1.747 | 2.044 | 1.638 |
| 20 | 1.826 | 1.318 | 2.434 | 2.074 | 2.091 | 2.168 | 1.747 |
| 25 | 2.168 | 1.673 | 2.355 | 2.184 | 2.574 | 2.122 | 1.809 |
| 50 | 2.309 | 1.879 | 3.4 | 2.325 | 3.183 | 2.808 | 2.839 |
| 75 | 2.605 | 2.279 | 3.229 | 2.871 | 3.744 | 3.167 | 3.151 |
| 100 | 2.964 | 2.611 | 3.932 | 4.024 | 4.774 | 4.306 | 4.134 |
| 150 | 2.822 | 3.503 | 4.711 | 6.708 | 5.429 | 4.945 | 5.663 |
| 200 | 4.899 | 4.486 | 5.865 | 10.359 | 8.144 | 6.942 | 8.487 |

We notice that the criterion function I2 has the least running time when compared to the other functions except when the number of clusters is 150.

## 8. Conclusion

We have shown that the Repeated bisection method performs the best with respect to clustering time when compared to the Direct and Agglomerative methods. We have analysed the quality of the clusters as we increased the number of clusters, for different criterion functions applied to the Repeated bisection method. We notice that there is no significant change in both Entropy and Purity beyond 50 clusters. Hence, we can conclude that even for large datasets, we need not increase the number of clusters beyond 50 while we can still maintain cluster quality.

We have also shown that among the Criterion functions, the I2 function performs the best with respect to clustering time.

Using clustering quality as a measure, we have shown that we can determine the maximum number of clusters for clustering documents.

## References

 [1] Jiawei Han, Micheline Kamber & Jian Pei "Data Mining Concepts and Techniques" second edition Morgan Kaufmann Publishers ISBN 13: 978-1-55860-901-3

[2] Pankaj Jajoo "Document clustering" IIT Kharagpur, Thesis,2008

 [3]A.K. Jain, M.N. Murty, P.J.Flynn, "Data Clustering: A Review", ACM Computing Surveys, Vol.31, No.3, September 1999.

[4] Ying Zhao & George Karypis , "Empirical and Theoretical Comparisons of Selected Criterion Functions for Document Clustering" , supported by NSF ACI-0133464, CCR-9972519, EIA-9986042, ACI-9982274, and by Army HPC Research Center.

[5] K. P. Soman, Shyam Diwakar, V. Ajay, "Insight Into Data Mining: Theory and Practice", 2006 by Prentice Hall of India Private Limited , ISBN-81-203-2897-3.

[6] Satya Chaitanya Sripada & Dr. M.Sreenivasa Rao, 'Comparison of purity and entropy of k-means clustering and fuzzy c means clustering' , Indian journal of computer science and engineering; Vol 2 no.3 June 2011; ISSN:0976-5166

[7]Van de Cruys, Tim " Mining for meaning: the extraction of lexico-semantic knowledge from text" Dissertation, Evaluation of cluster quality, chapter 6 , University of Groningen , 2010 .

[8]Anna Huang, University of Waikato, Hamilton, New Zealand, 'Similarity measures for Text Document Clustering' NZCSRSC 2008, April2008, Christ Church, New Zealand.

[9]   CLUTO-A Clustering Toolkit.    http://glaros.dtc.umn.edu/gkhome/views/cluto.